\documentclass{article}
\usepackage{graphics}
\usepackage{setspace}
\IfFileExists{url.sty}{\usepackage{url}}{\let\url=\verb}

\makeatletter

\providecommand{\LyX}{L\kern-.1667em\lower.25em\hbox{Y}\kern-.125emX\@}

\newenvironment{lyxlist}[1]
  {\begin{list}{}
    {\settowidth{\labelwidth}{#1}
     \setlength{\leftmargin}{\labelwidth}
     \addtolength{\leftmargin}{\labelsep}
     }}
  {\end{list}}
\newcommand{\lyxaddress}[1]{
  \par {\raggedright #1 
  \vspace{1.4em}
  \noindent\par}
}

\usepackage[T1]{fontenc}

\makeatletter

\date{}
\makeatother
\makeatother

\begin{document}

\title{Astrophysical Reaction Rates From Statistical Model Calculations}

\author{Thomas Rauscher and Friedrich-Karl Thielemann}

\maketitle

\lyxaddress{Departement f\"ur Physik und Astronomie, Universit\"at Basel, 
Klingelbergstr. 82, CH-4056 Basel, Schweiz (Switzerland)}

\lyxaddress{\copyright 2000 by Academic Press\\
The full article can be found in {\em Atomic Data Nuclear Data Tables}
{\bf 75} (2000) 1 and at \url{www.idealibrary.com}}

\begin{abstract}
\noindent Theoretical reaction rates in the temperature range \( 0.01\times
10^{9}\leq T({\rm K})\leq 10.0\times 10^{9} \)
are calculated in the statistical model (Hauser-Feshbach formalism)
for targets with \( 10\leq Z\leq 83 \) (Ne to Bi)
and for a mass range reaching the
neutron and proton driplines. Reactions considered are (n,\( \gamma  \)),
(n,p), (n,\( \alpha  \)), (p,\( \gamma  \)), (p,\( \alpha  \)), 
(\( \alpha  \),\( \gamma  \)),
and their inverse reactions. Reaction rates as a function of temperature
for thermally populated targets are given by analytic seven parameter fits.
To facilitate comparison with experimental rates, the stellar enhancement
factors are also tabulated. Two complete sets of rates have been
calculated, one of
which includes a phenomenological treatment of shell quenching for neutron-rich
nuclei. These extensive datasets are provided on-line as electronic
files, while a selected subset from one calculation is given as printed
tables. A summary of the theoretical inputs and advice on the use of the
provided tabulations is included. 
\end{abstract}
\tableofcontents{}

\section{Introduction}

Nuclear reaction rates are an essential ingredient for all investigations
of nucleosynthetic or energy generating processes in astrophysics. Highly
unstable nuclei are produced in such processes which again can become targets
for subsequent reactions. Cross sections and astrophysical reaction rates
for a large number of nuclei are required to perform complete network 
calculations
which take into account all possible reaction links and do not postulate
a priori simplifications. Despite concerted experimental efforts, most
of the involved nuclei are currently not accessible in the laboratory and
therefore theoretical models have to be invoked in order to predict reaction
rates.

In predictions of cross sections and reaction rates for astrophysical 
applications,
slightly different points are emphasized than in pure nuclear physics 
investigations.
Firstly, one is confined to the very low-energy region, from thermal energies
up to a few MeV. Secondly, since most of the ingredients for the calculations
are experimentally undetermined, one has to develop reliable phenomenological
models to predict these properties with an acceptable accuracy across the
nuclear chart. This task is made even harder by the lack of information
on specific properties, such as optical potentials for \( \alpha  \) particles
at the astrophysically relevant energies even for stable nuclei. Therefore,
one has to be satisfied with a somewhat more limited accuracy, as compared
to usual nuclear physics standards. The accuracy of the rates is estimated
to be within a factor of \( 1.5-2 \), with an even better average deviation,
e.g. of 1.4 for neutron capture. Considering the substantially larger 
uncertainties
in many astrophysical scenarios, this seems to be acceptable. Thus, the
real challenge is not the application of well-established models, but rather
to provide all the necessary ingredients in as reliable a way as possible,
also for nuclei where no such information is available. Many efforts have
been directed at addressing those problems and the current status of the
investigations make it worthwhile to publish a full set of theoretical
rates, intended to supersede early reaction rate tabulations 
\cite{A72,HWF76,WFH78,sar82,thi87,cowan91}.

For the majority of nuclear reactions in astrophysics, the statistical
model (Wolfenstein-Hauser-Feshbach approach) \cite{Wol51,HF52} can be
applied. This is appropriate provided the level density in the contributing
energy window around the peak of the projectile energy distribution is
sufficiently high to justify a statistical treatment. The critical level
density is usually estimated between 5 and 10 MeV\( ^{-1} \) \cite{RTK97}.
Furthermore, the compound nucleus picture will only dominate when the energy
of the incident particle is low enough (\( <20 \) MeV). While the latter
point is practically always satisfied in astrophysical environments, the
level density may fall below the critical value in certain nuclei lighter
than Fe, at shell closures, and for very neutron-rich or proton-rich isotopes
near the drip lines with correspondingly low separation energies. In these
cases, single resonances or direct capture contributions will become significant
and have to be treated individually. In this tabulation, we are not concerned
with such effects but rather give a full set of rates calculated in the
statistical model. However, the limits of its applicability will be discussed
(Sec.~\ref{sec:appl}).

In the following section, we concisely summarize the theoretical background
for easy reference as well as the nuclear properties used as input in the
calculations. This is followed by a section defining the reaction rates,
explaining the fitting procedure and giving more details on the tabulated
values. The paper is concluded by a summary and the rate tables.

\section{The Statistical Model}

\subsection{Theory}

The averaged transmission coefficients \( T \) comprise the central quantities
in statistical model calculations. They do not reflect a resonance behavior,
but rather describe absorption via an imaginary part in the (optical) 
nucleon-nucleus
potential~\cite{MW79}. This leads to the well-known expression

\noindent 
\begin{eqnarray}
\sigma ^{\mu \nu }(E_{ij}) & = & {{\pi \hbar ^{2}/(2\mu _{ij}E_{ij})}\over 
{(2J^{\mu }_{i}+1)(2J_{j}+1)}} \label{cslab}\\
 &  & \times \sum _{J,\pi }(2J+1){{T^{\mu }_{j}(E,J,\pi ,E^{\mu }_{i},
J^{\mu }_{i},\pi ^{\mu }_{i})T_{o}^{\nu }(E,J,\pi ,E^{\nu }_{m},J^{\nu }_{m},
\pi ^{\nu }_{m})}\over {T_{\mathrm{tot}}(E,J,\pi )}} \nonumber
\end{eqnarray}
for the cross section \( \sigma ^{\mu \nu } \) of the reaction 
\( i^{\mu }(j,o)m^{\nu } \)
from the target state \( i^{\mu } \) to the excited state \( m^{\nu } \)
of the final nucleus, with a center of mass energy \( E \)\( _{ij} \)
and reduced mass \( \mu _{ij} \). \( J \) denotes the spin, \( E \)
the corresponding excitation energy, and \( \pi  \)
the parity of excited states. When these properties are used without subscripts
they describe the compound nucleus; subscripts refer to states of the 
participating
nuclei in the reaction \( i^{\mu }(j,o)m^{\nu } \) and superscripts indicate
the specific excited states. The total transmission coefficient 
\( T_{\mathrm{tot}}=\sum _{\nu ,o}T_{o}^{\nu } \)
describes the transmission into all possible bound and unbound states \( \nu  \)
in all energetically accessible exit channels \( o \) (including the
entrance channel $i$). Experiments measure
\( \sigma ^{\mathrm{lab}}=\sum _{\nu }\sigma ^{0\nu }(E_{ij}) \), summed
over all excited states of the final nucleus, with the target in the ground
state. Target states \( \mu  \) in an astrophysical plasma of temperature
\( T^{*} \) are thermally populated and the astrophysical cross section
\( \sigma ^{*} \) is given by 
\begin{equation}
\label{csstar}
\sigma ^{*}(E_{ij})={\sum _{\mu }(2J^{\mu }_{i}+1)\exp (-E^{\mu }_{i}/kT^{*})
\sum _{\nu }\sigma ^{\mu \nu }(E_{ij})\over \sum _{\mu }(2J^{\mu }_{i}+1)
\exp (-E^{\mu }_{i}/kT^{*})}\quad ,
\end{equation}
$k$ being the Boltzmann constant.
The summation over \( \nu  \) replaces \( T_{o}^{\nu }(E,J,\pi ) \)
in Eq.~(\ref{cslab}) by the total transmission coefficient 
\begin{eqnarray}
T_{o}(E,J,\pi ) & = & \sum ^{\nu _{m}}_{\nu =0}T^{\nu }_{o}(E,J,\pi ,
E^{\nu }_{m},J^{\nu }_{m},\pi ^{\nu }_{m}) \label{tot}\\
 &  & +\int _{E^{\nu _{m}}_{m}}^{E-S_{m,o}}\sum _{J_{m},\pi _{m}}T_{o}(E,J,
\pi ,E_{m},J_{m},\pi _{m})\rho (E_{m},J_{m},\pi _{m})dE_{m}\quad .\nonumber
\end{eqnarray}
Here \( S_{m,o} \) is the channel separation energy, and the summation
over excited states above the highest experimentally known state \( \nu _{m} \)
is changed to an integration over the level density \( \rho  \). The summation
over target states \( \mu  \) in Eq.~(\ref{csstar}) has to be generalized
accordingly.

The important ingredients of statistical model calculations as indicated
in Eqs.~(\ref{cslab}) through (\ref{tot}) are the particle and 
$\gamma$-transmission
coefficients \( T \) and the level density of excited states \( \rho  \).
Therefore, the reliability of such calculations is determined by the accuracy
with which these components can be evaluated (often for unstable nuclei).
It is in these quantities that various statistical model calculations differ.
The reaction rates given in this paper are calculated with the code 
NON-SMOKER~\cite{rt98},
derived from the well-known SMOKER code~\cite{thi87}. (The code MOST~\cite{MOST}
is another code derived from SMOKER.) In the following we want to briefly
outline the methods utilized in that code to estimate those nuclear properties.
The challenge is in the goal to provide them in as reliable a way as possible,
also for unstable nuclei for which no experimental information is available.
Thus, global descriptions are employed which minimize the overall error
and are trusted to be reliable also far from stability.

\subsection{Transmission Coefficients}

The transition from an excited state in the compound nucleus \( (E,J,\pi ) \)
to the state \( (E^{\mu }_{i},J^{\mu }_{i},\pi ^{\mu }_{i}) \) in nucleus
\( i \) via the emission of a particle \( j \) is given by a summation
over all quantum mechanically allowed partial waves

\noindent 
\begin{equation}
\label{avtrans}
T^{\mu }_{j}(E,J,\pi ,E^{\mu }_{i},J^{\mu }_{i},\pi ^{\mu }_{i})=\sum _{l=|J-s|}
^{J+s}\sum _{s=|J^{\mu }_{i}-J_{j}|}^{J^{\mu }_{i}+J_{j}}T_{j_{ls}}
(E^{\mu }_{ij}).
\end{equation}
Here the angular momentum \( \vec{l} \) and the channel spin 
\( \vec{s}=\vec{J}_{j}+\vec{J}^{\mu }_{i} \)
couple to \( \vec{J}=\vec{l}+\vec{s} \). The transition energy in channel
\( j \) is \( E^{\mu }_{ij} \)=\( E-S_{j}-E^{\mu }_{i} \), where $S_j$ is
the channel separation energy.

The total transmission coefficients for this tabulation are then calculated
by applying Eq.~(\ref{tot}) and utilizing up to 19 experimentally known
excited states. The data are taken from \cite{toi96}, up to the first
level for which the spin assignment was not known. Ground state spin and
parities are known for many unstable nuclei. Far off stability, ground
state spins and parities are taken from \cite{mnk97}, if experimental
values are not available.

\subsubsection{Particle Transmission Coefficients}

The individual particle transmission coefficients \( T_{j_{ls}} \) are
calculated by solving the Schr\"{o}dinger equation with an optical potential
for the particle-nucleus interaction. We employ the optical potential for
neutrons and protons given by~\cite{JLM77}, based on microscopic infinite
nuclear matter calculations for a given density, applied with a local density
approximation. It includes corrections of the imaginary part~\cite{F81,M82}. 

The optical potential for \( \alpha  \) particles from \cite{Mc66} was
shown to be quite accurate for a wide range of nuclei and is used in this
work. However, it was realized \cite{morau97,som98} that for heavily charged
nuclei a more sophisticated potential had to be adopted at the comparatively
low energies of astrophysical interest. 
Promising is the folding approach~\cite{SL79},
with a parameterized mass- and energy-dependence of the real volume 
integral~\cite{rau98a}.
Microscopic and deformation information should be considered in the 
parametrization
of the imaginary potential~\cite{rau98b}. However, due to the scarcity
of experimental data, the potential parameters can as yet only be extracted for
a limited mass and energy range. The optical $\alpha$+nucleus potential is
likely to introduce the largest uncertainties in the charged particle rates
presented here. Further experimental work is clearly necessary and most
welcome.

Deformed nuclei are treated by using an effective spherical potential of
equal volume, based on averaging the deformed potential over all possible
angles between the incoming particle and the orientation of the deformed
nucleus.

\subsubsection{Radiative transmission coefficients}

At least the dominant $\gamma$-transitions (E1 and M1) have to be included
in the calculation of the total photon width. The smaller, and therefore
less important, M1 transitions are treated, as usual, in the simple single
particle approach (\( T\propto E^{3} \)~\cite{BW52}), as also discussed
in \cite{HWF76}. The E1 transitions are calculated on the basis of the
Lorentzian representation of the Giant Dipole Resonance (GDR). Within this
model, the E1 transmission coefficient for the transition emitting a photon
of energy \( E_{\gamma } \) in a compound nucleus \( ^{A}_{N}Z \) is given by
\begin{equation}
\label{GDR}
T_{E1}(E_{\gamma })={8\over 3}{NZ\over A}{e^{2}\over \hbar c}{{1+\chi }\over 
Mc^{2}}\sum _{i=1}^{2}{i\over 3}{{\Gamma _{G,i}E^{4}_{\gamma }}\over 
{(E_{\gamma }^{2}-E^{2}_{G,i})^{2}+\Gamma ^{2}_{G,i}E^{2}_{\gamma }}}\quad .
\end{equation}
Here, $M$ is the proton mass, \( \chi (=0.2) \) accounts for the 
neutron-proton exchange contribution~\cite{LS89},
and the summation over \( i \) includes two terms which correspond to
the split of the GDR in statically deformed nuclei, with oscillations along
(\( i=1 \)) and perpendicular (\( i=2 \)) to the axis of rotational symmetry.
Many microscopic and macroscopic models have been devoted to the calculation
of the GDR energies (\( E_{G} \)) and widths (\( \Gamma _{G} \)). Here,
the (hydrodynamic) droplet model approach~\cite{My77} is used for \( E_{G} \),
which gives an excellent fit to the GDR energies and can also predict the
split of the resonance for deformed nuclei, when making use of the deformation,
calculated within the droplet model. In that case, the two resonance energies
are related to the mean value calculated by the relations~\cite{D58} 
\( E_{G,1}+2E_{G,2}=3E_{G} \),
\( E_{G,2}/E_{G,1}=0.911\eta +0.089 \). \( \eta  \) is the ratio of the
diameter along the nuclear symmetry axis to the diameter perpendicular
to it, and is obtained from the experimentally known deformation or mass
model predictions. For the width \( \Gamma _{G} \) of the GDR the description
of~\cite{TA83} is used, which applies to spherical and deformed nuclei
and can be described as a superposition of a macroscopic width due to the
viscosity of the nuclear fluid and a coupling to quadrupole surface vibrations
of the nucleus (see also \cite{cowan91}). 

Direct application of Eq.~(\ref{GDR}) would overestimate the radiation width
by about 30\% (see e.g. \cite{McC81,Ram81}). This is due to the fact that,
for low energy \( \gamma  \)-transitions, the Lorentz curve is suppressed
and the GDR width increases with excitation energy (e.g. \cite{kad82,kop90}).
To account for these deficiencies, various treatments of an energy-dependent
width have been suggested. We use the form~\cite{McC81} 
\begin{equation}
\Gamma _{G}\left( E_{\gamma }\right) =\Gamma 
_{G}\sqrt{\frac{E_{\gamma }}{E_{G}}\: .}
\end{equation}

Another effect has to be taken into account for certain \( \alpha  \)-capture
reactions. Because of isospin selection rules, \( \gamma  \)-transitions
between levels with isospin \( I=0 \) are forbidden. This leads to a suppression
of the cross section for (\( \alpha  \),\( \gamma  \)) reactions on 
self-conjugate
(\( N=Z \)) targets, due to isospin conservation. A suppression could
also be found for capture reactions leading into self-conjugate nuclei,
although somewhat less pronounced because \( I=1 \) states can be populated
according to the isospin coupling coefficients. The suppression is usually
treated as a suppression of the \( \gamma  \)-width. In previous rate
tabulations it was either neglected~\cite{thi87} or accounted for in a
phenomenological way by dividing the \( \gamma  \)-width by quite arbitrary
factors~\cite{HWF76,WFH78}. In the code NON-SMOKER the appropriate 
\( \gamma  \)-widths
are obtained by explicitly accounting for isospin mixing and suppression
of the appropriate \( \gamma  \)-transitions~\cite{rt98}. A detailed
account of the procedure can be found in \cite{RGW99}.

\subsubsection{Width fluctuation corrections}

In addition to the ingredients required for Eq.~(\ref{cslab}), like the
transmission coefficients for particles and photons, width fluctuation
corrections \( W(j,o,J,\pi ) \) have to be employed. They define the correlation
factors with which all partial channels for an incoming particle \( j \)
and outgoing particle \( o \), passing through the excited state 
\( (E,J,\pi ) \),
have to be multiplied. This takes into account that the decay of the state
is not fully statistical, but some memory of the way of formation is retained
and influences the available decay choices. The major effect is elastic
scattering, the incoming particle can be immediately re-emitted before
the nucleus equilibrates. Once the particle is absorbed and not re-emitted
in the very first (pre-compound) step, the equilibration is very likely.
This corresponds to enhancing the elastic channel by a factor \( W_{j} \).
In order to conserve the total cross section, the individual transmission
coefficients in the outgoing channels have to be renormalized to 
\( T_{j}^{\prime } \).
The total cross section is proportional to \( T_{j} \) and, when summing
over the elastic channel (\( W_{j}T_{j}^{\prime } \)) and all outgoing
channels (\( T^{\prime }_{tot}-T^{\prime }_{j} \)), one obtains the condition
\( T_{j} \)=\( T_{j}^{\prime }(W_{j}T_{j}^{\prime }/T^{\prime }_{tot})+
T^{\prime }_{j}(T^{\prime }_{tot}-T^{\prime }_{j})/T^{\prime }_{tot} \)
\cite{V84}. We can (almost) solve for \( T^{\prime }_{j} \)
\begin{equation}
\label{widthcorr}
T^{\prime }_{j}={T_{j}\over 1+T^{\prime }_{j}(W_{j}-1)/T^{\prime }_{tot}}\quad .
\end{equation}
This requires an iterative solution for \( T^{\prime } \) (starting in
the first iteration with \( T_{j} \) and \( T_{tot} \)), which converges
rapidly. The enhancement factor \( W_{j} \) has to be known in order to apply
Eq.~(\ref{widthcorr}). 
A general expression in closed form was derived~\cite{V84},
but is computationally expensive to use. A fit to results from Monte Carlo
calculations gave~\cite{T74}
\begin{equation}
\label{newcorr}
W_{j}=1+{2\over 1+T_{j}^{1/2}}\quad .
\end{equation}

For a general discussion of approximation methods see \cite{GH92,EP93}.
Eqs.~(\ref{widthcorr}) and (\ref{newcorr}) redefine the transmission 
coefficients
of Eq.~(\ref{cslab}) in such a manner that the total width is redistributed
by enhancing the elastic channel and weak channels over the dominant one.
Cross sections near threshold energies of new channel openings, where very
different channel strengths exist, can only be described correctly when
taking width fluctuation corrections into account. The width fluctuation
corrections of~\cite{T74} are only an approximation to the correct treatment.
However, it was shown that they are quite adequate~\cite{TZL86}.

\subsection{Level Densities}

\label{sec:levden}Until recently, the nuclear level density has given
rise to the largest uncertainties in the description of nuclear reactions
in the statistical model~\cite{HWF76,thi87,cowan91}. Implemented in the
NON-SMOKER code is a recently improved treatment~\cite{RTK97}. It is based
on a shifted Fermi-gas formalism~\cite{gil65} with an energy-dependent
level density parameter \( a \) together with microscopic corrections
from nuclear mass models. This leads to improved fits to known level densities
in a wide range of masses~\cite{RTK97}. More sophisticated Monte Carlo
shell model calculations~\cite{dean95}, as well as combinatorial approaches
(see e.g.~\cite{paar97}), have shown excellent agreement with this 
phenomenological
approach and justified the application of the Fermi-gas description at
and above the neutron separation energy. 

An in-depth description of the model and its application to astrophysical
problems can be found in~\cite{RTK97}. Here, we only want to briefly summarize
the inputs used for calculating the rates presented in this tabulation.
It should be noted that we applied our description throughout the nuclear
chart, without relying on experimental level density parameters in specific
cases as has been done before \cite{HWF76,WFH78,MOST}. This may lead to
locally slightly larger deviations from experiment but it improves the
reliability when extrapolating to unknown isotopes.

The microscopic correction and the pairing corrections comprise crucial
inputs for the level density formalism used here (see \cite{RTK97} for
details). They can be extracted from mass models. There is a choice of
several mass models in the NON-SMOKER code. The Finite Range Droplet Model
(FRDM)~\cite{frdm} and an extended Thomas-Fermi approach with Strutinski
Integral (ETFSI-Q)~\cite{etfsiq} have been chosen for the reaction rate
calculations in this work. It has to be emphasized that experimental mass
values \cite{audi95} were included where possible. This is straightforward
for the separation energies which were calculated from the mass differences;
it was ensured that either only experimental or theoretical values were
used in the differences, thus avoiding unphysical breaks at transition
points from experiment to theory. The microscopic corrections were obtained
by subtracting the well-defined spherical macroscopic (droplet) term of
the FRDM from the total mass energy derived from experiment, from the FRDM
or from ETFSI-Q, respectively (cf.~Eq.~(17) in \cite{RTK97}). The validity
of the resulting rates is discussed in Sec.~\ref{sec:fits}. Rates based on
other mass models can be obtained from the authors on request or on-line
(see Sec.~\ref{sec:summary}).

The shifted Fermi-gas approach diverges for \( U=E-\delta =0 \) (i.e.\ 
\( E=\delta  \),
if \( \delta  \) is a positive backshift, with \( E \) being the excitation
energy and \( \delta  \) being an energy shift due to pairing corrections).
In order to obtain the correct behavior at very low excitation energies,
the Fermi-gas description can be combined with the constant temperature
formula (\cite{gil65}; \cite{GH92} and references therein) 
\begin{equation}
\label{ctemp}
\rho (U)\propto {\exp (U/T_{\mathrm{nucl}})\over T_{\mathrm{nucl}}}\quad .
\end{equation}
 The two formulations are matched by a tangential fit determining the nuclear
temperature \( T_{\mathrm{nucl}} \).

\subsection{Applicability of the Statistical Model}

\label{sec:appl}The statistical model can be applied provided that the
use of averaged transmission coefficients (Eq.~(\ref{avtrans})) is permitted.
This will be the case for high level densities with completely overlapping
resonances, typical for the compound nucleus reaction mechanism. For light
nuclei, decreasing particle separation energies or at shell closures, level
densities will eventually become too low for the application of the statistical
model at astrophysical temperatures. In those cases, single resonances
and contributions from the direct reaction mechanism have to be taken into
account \cite{raubi98}. Based on the level density description outlined
in Sec.~\ref{sec:levden}, a quantitative criterion for the applicability
was derived recently~\cite{RTK97}. In the present work we give tables
of all reaction rates regardless of applicability but specify the allowed
temperature range in the tables. The estimate is quite conservative and
thus the rates can still be accurate slightly below the given lower limits
of the temperature.

\section{Astrophysical Reaction Rates}

\subsection{Definition}

The nuclear reaction rate per particle pair at a given stellar temperature
\( T^{*} \) is determined by folding the reaction cross section 
\( \sigma  \)\( ^{*}(E) \) from Eq.~(\ref{csstar})
with the Maxwell-Boltzmann velocity distribution of the 
projectiles~\cite{fowler}
\begin{equation}
\label{rate}
\left\langle \sigma ^{*}v\right\rangle =\left\langle \sigma v\right\rangle 
^{*}=\left( \frac{8}{\pi \mu }\right) ^{1/2}\frac{1}{\left( kT^{*}\right) 
^{3/2}}\int\limits _{0}^{\infty }\sigma ^{*}(E)E\exp \left( 
-\frac{E}{kT^{*}}\right) dE\quad .
\end{equation}
It has to be emphasized that only the use of the stellar cross section
\( \sigma  \)\( ^{*} \) (Eq.~(\ref{csstar})) yields a reaction rate with
the desired behavior that the inverse reaction can be calculated by using
detailed balance. Therefore, laboratory rates -- which only measure 
\( \sigma  \)\( ^{\mathrm{lab}}=\sum _{\nu }\sigma ^{0\nu } \),
i.e. the cross section with the target being in the ground state -- should
always be measured in the direction that is least affected by excited target
states. This is usually the reaction with positive \( Q \)-value (exoergic
reaction). For astrophysical applications, such rates have to be corrected
for the stellar enhancement effect due to the thermal excitation of the
target \cite{A72}. The stellar enhancement factors (SEF) \( f^{*} \)
are defined by 
\begin{equation}
\label{eq:sef}
f^{*}=\frac{\sigma ^{*}}{\sigma ^{\mathrm{lab}}}\quad .
\end{equation}
The values of \( f^{*} \)for a range of temperatures for nuclei close
to stability are given in the tables. Stellar enhancement factors for neutron
capture reactions and a discussion of the involved uncertainties can also
be found in a recent compilation of neutron cross sections for the \( s \)
process \cite{bao99}.

\subsection{Partition functions and reverse rates}

The temperature-dependent partition function \( G(T^{*}) \) normalized
to the ground state target spin \( J_{i}^{0} \) is defined as \cite{fowler67}
\begin{eqnarray}
(2J_{i}^{0}+1)G(T^{*})= &  & \sum _{\mu =0}^{\mu _{m}}(2J_{i}^{\mu }+1)
e^{-E_{i}^{\mu }/kT^{*}} \label{eq:partfunc}\\
 &  & +\int\limits _{E_{i}^{\mu _{m}}}^{E_{i}^{\mathrm{max}}}\sum _{J^{\mu },
\pi ^{\mu }}(2J^{\mu }+1)
e^{-\epsilon/kT^{*}}\rho (\epsilon ,J^{\mu },\pi ^{\mu })d\epsilon 
\quad ,\nonumber 
\end{eqnarray}
with $\rho$ being the level density and \( \mu  \)\( _{m} \) the
last included experimentally known state. The included experimental levels
were the same as for the calculation of the transmission coefficients 
(Eq.~(\ref{tot})).
For the temperature range considered here, the maximum energy 
\( E_{i}^{\mathrm{max}} \)
above which there are no more significant contributions to the partition
function is of the order of \( 20-30 \) MeV. With that definition, the
stellar reaction rate \( \left\langle \sigma _{i}v\right\rangle ^{*} \)
for a reaction with particles in all channels is related to the rate of
the reverse reaction \( \left\langle \sigma _{m}v\right\rangle ^{*} \)
by
\begin{equation}
\label{invpart}
N_{A}\left\langle \sigma _{m}v\right\rangle ^{*}=\left( 
\frac{A_{i}A_{j}}{A_{o}A_{m}}\right) 
^{3/2}\frac{(2J_{i}+1)(2J_{j}+1)}{(2J_{o}+1)(2J_{m}+1)}\frac{G_{i}
(T^{*})}{G_{m}(T^{*})}e^{-Q/kT^{*}}N_{A}\left\langle 
\sigma _{i}v\right\rangle ^{*}\quad ,
\end{equation}
where $N_A$ is Avogadro's number, $J$ and $A$ are spins and masses \( A \) 
(in atomic mass units $u$) of the particles involved in the reaction
$i$($j$,$o$)$m$, $Q$ is the reaction $Q$-value.
To calculate
photodisintegration rates from capture rates the appropriate relation
is
\begin{equation}
\label{invphot}
\lambda _{\gamma }=\left( \frac{A_{i}A_{j}}{A_{m}}\right) 
^{3/2}\frac{(2J_{i}+1)(2J_{j}+1)}{(2J_{m}+1)}\frac{G_{i}(T^{*})}
{G_{m}(T^{*})}\left( T^{*}\right) ^{3/2}Fe^{-Q/kT^{*}}N_{A}\left\langle 
\sigma _{i}v\right\rangle ^{*}\quad .
\end{equation}
For $\lambda_\gamma$ in s$^{-1}$ and using the usual practical units,
i.e. temperatures $T_9=T^*/10^9$ K and 
\( N_{A}\left\langle \sigma v\right\rangle ^{*} \)
in cm\( ^{3} \) s\( ^{-1} \) mole\( ^{-1} \), one obtains
\begin{equation}
\left(T^*\right)^{3/2}F=\left( \frac{ukT^{*}}{2\pi \hbar ^{2}}\right) 
^{3/2}\frac{1}{N_A}=T_9^{3/2} 9.8685\times 10^9 {\rm mole\;
cm}^{-3}\quad .
\end{equation}
The numerical factor $F$ as well as the spin and mass factors are already 
accounted for in
the parameter \( a_{0}^{\mathrm{rev}} \) which is tabulated. See 
Sec.~\ref{sec:revpar}
for further details.

In the tabulated rates, the thermal effects are already considered. Statistical
model calculations not including full photon cascades may also be prone
to some error arising from the decay of unbound particle states in reactions
with negative \( Q \)-values (endoergic reactions). Furthermore, due to
the exponential dependence of the inverse rate on the \( Q \)-value 
(Eqs.~(\ref{invpart})
and (\ref{invphot})), inaccuracies in the rate would be strongly enhanced
when computing the exoergic rate from the endoergic one. In order to minimize
the error, reactions are always calculated in the exoergic direction (with
the exceptions of capture reactions, where photodisintegration is always
treated as the reverse reaction regardless of $Q$-value) and 
detailed balance is applied to
obtain the values for the endoergic reaction. This treatment has the additional
advantage that it ensures consistent values for forward and reverse reactions,
which is essential for application in astrophysical nuclear reaction networks.
To calculate the actual (endoergic) reaction rate, those fits have to be
multiplied by the ratio of the partition functions of the final nucleus
and the target \( G_{i}/G_{m} \) at the appropriate temperature 
(see Sec.~\ref{sec:revpar}
and Sec.~\ref{sec:examples}). For that purpose, partition functions are
tabulated separately.

\subsection{Analytic reaction rate fits\label{sec:fits}}

\subsubsection{Parametrization}

Reaction rates have been calculated for a temperature grid of 24
temperatures: $T_9$=0.1, 0.15, 0.2, 0.3, 0.4, 0.5, 0.6, 0.7, 0.8, 0.9,
1.0, 1.5, 2.0, 2.5, 3.0, 3.5, 4.0, 4.5, 5.0, 6.0, 7.0, 8.0, 9.0, 10.0.
For easy application in astrophysical investigations, all reaction types
((n,\( \gamma  \)), (n,p), (n,\( \alpha  \)), (p,\( \gamma  \)), (p,n),
(p,\( \alpha  \)), (\( \alpha  \),\( \gamma  \)), (\( \alpha  \),n),
(\( \alpha  \),p), (\( \gamma  \),n), (\( \gamma  \),p), 
(\( \gamma  \),\( \alpha  \)))
are fitted with the same parametrization
\begin{eqnarray}
\left. \begin{array}{c}
N_{A}\left\langle \sigma v\right\rangle ^{*}\\
\lambda_\gamma
\end{array} \right\}&=&\exp \left( a_{0}+a_{1}T_{9}^{-1}+a_{2}T_{9}^{-1/3}
+a_{3}T_{9}^{1/3}+a_{4}T_{9} \right. \nonumber\\
 & &\left. +a_{5}T_{9}^{5/3}+a_{6}\ln T_{9}\right) \quad ,
\label{fitpar}
\end{eqnarray}
with the seven open parameters \( a_{0}-a_{6} \) and the stellar temperature
\( T_{9} \) given in 10\( ^{9} \) K. This parametrization proves to be
flexible enough to accommodate the different temperature dependencies of
the various reaction types across the fitted temperature range of 
\( 0.01\leq T_{9}\leq 10. \)
Parametrizations of the present rates in the form used in \cite{HWF76}
and others can be obtained from the authors on request.

\subsubsection{Parameters for the reverse rates}

\label{sec:revpar}The parameters for the reverse rates are not given explicitly
but can easily be computed from the information in the tables. To calculate
the reverse rate of the reaction \( i(j,o)m \), i.e. the reaction \( m(o,j)i \),
Eq.~(\ref{fitpar}) is employed and the seven parameters 
\( a_{0}^{\mathrm{rev}}-a_{6}^{\mathrm{rev}} \)
for the reverse rate are determined as follows: 
\begin{eqnarray}
a_{0}^{\mathrm{rev}} & = & a_{0}^{\mathrm{rev}},\quad \mathrm{as}\; 
\mathrm{tabulated} \nonumber\\
a_{1}^{\mathrm{rev}} & = & a_{1}-11.6045Q \nonumber\\
a_{2}^{\mathrm{rev}} & = & a_{2} \nonumber\\
a_{3}^{\mathrm{rev}} & = & a_{3} \nonumber\\
a_{4}^{\mathrm{rev}} & = & a_{4} \label{revcoff}\\
a_{5}^{\mathrm{rev}} & = & a_{5} \nonumber\\
a_{6}^{\mathrm{rev}} & = & \left\{ \begin{array}{lr}
a_{6}+1.5 & \quad (\mathrm{i})\\
a_{6} & (\mathrm{ii})
\end{array}\right. \nonumber
\end{eqnarray}
The above relations are derived from Eqs.~(\ref{invpart}) and (\ref{invphot}),
using Eq.~(\ref{fitpar}) and taking the logarithms on both sides. For
the coefficient $a_6^{\rm rev}$,
case (i) applies when calculating a photodisintegration rate from a capture
rate, case (ii) for all other rates. Finally, for the reverse reaction
case, {\em the value found by application
of Eqs.~(\ref{fitpar}) and (\ref{revcoff}) has to be multiplied by the 
ratio of the partition
functions for residual and target nucleus \( G_{i}/G_{m} \)}. Examples
are shown in Sec.~\ref{sec:examples}.

\subsubsection{Fit accuracy}

The flexibility of the fitting function makes it prone to numerical problems
outside the calculated range at low temperatures, where the rates should
be close to zero. In some cases they tend to diverge strongly. This difficulty
can be avoided by providing fit data at low temperatures additionally to
the calculated values by appropriately extrapolating the rates to lower
temperatures. This is achieved by either assuming s-wave capture for 
\( T_{9}<0.1 \)
for exoergic neutron capture reactions 
(Maxwellian averaged capture cross sections
in the energy range \( 5\leq E\leq 100 \) keV for targets along the line
of stability can be found in another compilation \cite{bao99}) or by considering
proper Coulomb barrier penetration factors in the charged particle channels.
Thus, both accuracy and flexibility can be ensured within a single 
parametrization.
However, it has to be emphasized that \emph{the given parameterization
is only valid within the temperature range of} \( 0.01\leq T_{9}\leq 10. \),
although many fits will show ``proper'' behavior down to lower
temperature. 
Caution is advised when using derived ($\gamma$,p) and
($\gamma$,n) rates at the proton dripline (see below).
For all cases, it is recommended to use the fits only down to
the temperature \( T_{\rm low}^{\rm fit} \) given in
the table. The temperatures of the validity
of the fits are given in the tables for each reaction, to emphasize the
importance of the given fit ranges.

As a measure of the accuracy of a given fit, the quantity \( \zeta  \)
is quoted in the tables. It is defined by
\begin{equation}
\label{eq:accur}
\zeta =\frac{1}{24}\sum _{i=1}^{24}\left( \frac{r_{i}-f_{i}}{f_{i}}\right) ^{2}\quad ,
\end{equation}
with \( r \) being the original rate value as calculated at each of the
24 temperatures $T_9$= 0.1, 0.15 \dots  10.0, and \( f \) the
rate calculated from the fit at these temperatures. Contributions with
$r<10^{-20}$ cm$^3$ s$^{-1}$ mole$^{-1}$ are neglected as lower accuracy at
at low rates is inconsequential. Note that while a small value of
$\zeta$ is indicative of an accurate fit over the entire temperature
range, large $\zeta$ generally signify deviations of the calculated from
the fitted rate at the lowest temperatures only.

The fit parameters are tabulated regardless of the validity of the statistical
model of nuclear reactions in the given temperature range 
(see Sec.~\ref{sec:appl}).
The estimated lower temperature limit of the validity of the statistical
model, $T_{\rm low}^{\rm HF}$ is given separately
for each rate in the tables. Below that limit the calculation of the rate
by means of the statistical model may not be justified, although the fit
to the calculated rate will still be accurate. At temperatures below the
applicability limit, rates may be over-estimated and should be compared
to calculations considering single resonance and direct reaction contributions.
Especially close to the driplines, fits of reactions with low $Q$-value
cannot be applied at low temperatures.
Although the fit may be valid, it should not
be used at low temperature because the statistical model will not be
applicable anymore.

\subsubsection{Computed rate sets}

Two different sets of rates have been calculated. They differ
in the mass model used, which enters into the computation of the separation
energies and \( Q \)-values as well as into the microscopic input to the
level density calculation (see Sec.~\ref{sec:levden}). One set was calculated
employing the well-known FRDM mass model \cite{frdm}, which excellently
reproduces masses and other ground state properties of nuclei close to
stability. It is also the most comprehensive set across the nuclear chart. 

Recently, it was suggested that so-called shell quenching effects may arise
for neutron-rich nuclei far off stability \cite{kratz93,nuc_phys_thie}.
Fully microscopic calculations \cite{chen95,doba96} and experimental data
\cite{wern94,hoff96,kaut96,zhang97} indicate the weakening of nuclear
shell gaps for neutron-rich nuclei. In the absence of microscopic calculations
of the required nuclear properties for the whole nuclear chart, it is possible
to phenomenologically include such quenching into existing mass formulae.
This has been done in the ETFSI-Q model \cite{etfsiq}, based on the ETFSI-1
mass model \cite{aboussir92,etfsi1}. However, it does not cover the full
range of isotopes. Therefore, we provide the alternative sets of reaction
rates obtained with the two mass models, so that the rates based on the
FRDM can be used for large-scale studies and close to stability, and the
rates based on ETFSI-Q for investigations concerning neutron-rich unstable
isotopes and the \( r \) process. It has to be noted that one should refrain
from mixing rates from the two sets as this will lead to inconsistencies
and artificial effects in the results.

\subsubsection{Mass ranges of the tabulated fits}

Due to the extensive number of nuclear reactions in the considered mass
range, we have to limit the printed version of our reaction rate fits.
Full rate libraries both for reactions calculated with the FRDM and with
the ETFSI-Q (as well as ETFSI-1) mass model can be obtained on-line or
from the authors on request (see Sec.~\ref{sec:summary}). In the printed
version, only the FRDM set is given and no capture rates for reactions
with negative
\( Q \) value are shown.

The full electronic versions of the tables available on-line include all 
reactions in the range \( 10\leq Z \leq 83 \) (FRDM) and $24 \leq Z \leq 83$
(ETFSI-Q). This amounts to 5369 (FRDM) and 4628 (ETFSI-Q) involved nuclei.
The isotope ranges for which rate
fits are available
are given in Table A; for the FRDM, the mass range is also indicated by
the heavy lines in Tables IA-IC. Rate fits are given for all n-, p-, and
$\alpha$-capture reactions and for those (n,p), (n,$\alpha$), (p,n),
(p,$\alpha$), ($\alpha$,n), and ($\alpha$,p) reactions having positive
$Q$-value. Reverse rates are not given explicitly but can be computed by
a two-step procedure as described in Secs.\ \ref{sec:revpar} and 
\ref{sec:examples}.
The stellar enhancement factors close to stability as well as the partition
functions for all isotopes are given for a temperature grid of 24 temperatures:
\( T_{9} \) = 0.1, 0.15, 0.2, 0.3, 0.4, 0.5, 0.6, 0.7, 0.8, 0.9, 1.0, 1.5,
2.0, 2.5, 3.0, 3.5, 4.0, 4.5, 5.0, 6.0, 7.0, 8.0, 9.0, 10.0 .

The printed Tables II\( - \)IV contain the following (calculated with
the FRDM mass model):

\begin{itemize}
\item (n,\( \gamma  \)) rate fits from stability to close to the neutron
dripline in the range $10\leq Z\leq 49$, and to $N=Z+59$ for 
$50\leq Z\leq 83$ (Table II).
\item (n,p) and (n,\( \alpha  \)) rate fits around stability in the range 
\( 10 \leq Z \leq 83 \)
(Table II).
\item (p,\( \gamma  \)), (p,n), (p,\( \alpha  \)), (\( \alpha  \),\( 
\gamma  \)),
(\( \alpha  \),n), and (\( \alpha  \),p) rate fits from proton-rich
nuclei to stability in the range \( 10\leq Z \leq 50 \) (Tables III and IV).
\item Reverse (endoergic) rates are not given explicitly but can be computed
with the help of partition functions
from the information given in the tables (see Secs.~\ref{sec:revpar},
\ref{sec:examples}).
\item Stellar enhancement factors \( f^{*} \) at selected 16 temperatures are
only quoted close to the valley of stability (Tables II\( - \)IV).
\item Partition functions for all involved isotopes are given at selected 20
temperatures in the range $0.1\leq T_9 \leq 10$ (Table V).
\end{itemize}
An overview of the provided rates is given in Tables IA\( - \)IC, which
show in detail for which n-, p-, and $\alpha$-induced reactions rate fit
parameters are available in Tables II$-$IV, respectively.

\subsubsection{Examples of use of tables}

\label{sec:examples}This section is intended to help with interpreting
the information given in the tables. We give two examples for calculating
the reaction rate for a given reaction and its inverse reaction at a temperature
of \( T_{9}=2.0 \).

The first example is the capture reaction \( ^{35} \)Ar(p,\( 
\gamma  \))\( ^{36} \)K.
From Table III one finds a \( Q \)-value of \( Q=1.666 \) MeV and the
parameters \( a_{0}= \)128.39, \( a_{1}=-4.0033 \), \( a_{2}=137.67 \),
\( a_{3}=-276.87 \), \( a_{4}= \)17.691, \( a_{5}=-1.0728 \), 
\( a_{6}= \)123.68.
With the help of Eq.~(\ref{fitpar}) one calculates \( N_{A}\left\langle 
\sigma v\right\rangle ^{*}=92.5 \)
cm\( ^{3} \)s\( ^{-1} \)mole\( ^{-1} \) at \( T_{9}=2.0 \). Because
both \( T_{\mathrm{low}}^{\mathrm{HF}} \) and \( 
T_{\mathrm{low}}^{\mathrm{fit}} \)
are considerably smaller than our temperature \( T_{9} \), it is safe
to assume that the statistical model is applicable and the fit to the rate
is valid. In order to obtain the value for the reverse rate, one first
has to determine the parameter values in the given parametrization. The
parameter \( a_{0}^{\mathrm{rev}}=151.83 \) is given in the table. The
remaining parameters are derived according to Eq.~(\ref{revcoff})
for a photodisintegration rate. 
This yields \( a_{1}^{\mathrm{rev}}=-23.3364 \),
\( a_{6}^{\mathrm{rev}}=125.18 \); all other parameters assume the same
value as for the forward reaction. Using those, Eq.~(\ref{fitpar}) gives
a value of \( \lambda_\gamma  \)\( ^{\prime }= \)2.5\( \times  \)10\( 
^{8} \)
s\( ^{-1} \). This has to be multiplied by the ratio of the partition
functions in order to obtain the valid rate factor for 
\( ^{36} \)K(\( \gamma  \),p)\( ^{35} \)Ar:
\[
\lambda_\gamma =\lambda_\gamma ^{\prime }\frac{G_{^{35}\mathrm{Ar}}}{
G_{^{36}\mathrm{K}}}=\lambda_\gamma 
^{\prime }\frac{1.001}{1.203}=2.1\times 10^{8}\quad \mathrm{s}^{-1}.\]
The values of the partition functions at \( T_{9}=2.0 \) were taken from
Table V. Note that for capture rates the procedure is always
the same as described
above regardless of whether it is an exoergic or an endoergic reaction.

The second example we consider is the reaction
\( ^{34} \)S(\( \alpha  \),n)\( ^{37} \)Ar,
again at \( T_{9}=2.0 \). It is not to be found in Table IV because of
its negative \( Q \)-value. Therefore, one has to rely on Table II to
calculate this reaction as the inverse reaction of 
\( ^{37} \)Ar(n,\( \alpha  \))\( ^{34} \)S.
The parameters found in Table II are \( a_{0}= \)20.072, \( a_{1}=-0.019613 \),
\( a_{2}=1.8224 \), \( a_{3}=-4.759 \), \( a_{4}= \)0.56437, 
\( a_{5}=-0.033893 \),
\( a_{6}= \)1.7801, and \( Q=4.63 \) MeV. Again, both \( 
T_{\mathrm{low}}^{\mathrm{HF}} \)
and \( T_{\mathrm{low}}^{\mathrm{fit}} \) are lower than the temperature
of interest. The rate \( N_{A}\left\langle \sigma v\right\rangle ^{*}=5.2\times
10^{7} \)
cm\( ^{3} \)s\( ^{-1} \)mole\( ^{-1} \) for \( ^{37} \)Ar(n,\( \alpha  \))\( 
^{34} \)S
results from the direct application of Eq.~(\ref{fitpar}). In order to calculate
the rate of \( ^{34} \)S(\( \alpha  \),n)\( ^{37} \)Ar, the parameters
are determined by application of Eq.~(\ref{revcoff}). This
yields the value \( a_{1}^{\mathrm{rev}}=-53.748448 \). 
The value of \( a_{0}^{\mathrm{rev}}=20.199 \)
is taken from the table and all other parameters remain the same as for
the forward reaction. With Eq.~(\ref{fitpar}) one arrives at the rate value 
\( r^{\prime}=1.272\times 10^{-4} \)
cm\( ^{3} \)s\( ^{-1} \)mole\( ^{-1} \) at \( T_{9}=2.0 \). Multiplying
this by the appropriate ratio of partition functions taken from Table V
yields the final result
\[
N_{A}\left\langle \sigma v\right\rangle ^{*}=r^{\prime}\frac{G_{
^{37}\mathrm{Ar}}}{G_{^{34}S}}=r^{\prime}\frac{1.0}{1.0}=1.27\times 
10^{-4}\quad \mathrm{cm}^{3}\mathrm{s}^{-1}\mathrm{mole}^{-1}.\]

\section{Summary}

\label{sec:summary}Thermonuclear reaction rates for neutron-, proton-
and \( \alpha  \)-induced reactions and their inverses have been calculated
in the statistical model. All rates from the proton dripline to the
neutron dripline for $10 \leq Z \leq 83$
(Ne to Bi) have been fitted to a unique function with seven free parameters.
Tables of these parameters are provided on-line 
for two sets of rates, calculated
with input from two different mass models. Furthermore, the stellar enhancement
factors are given in order to facilitate comparison with experimental ground
state rates. A printed subset of the on-line tables for the FRDM
presented here shows fit parameters for (n,$\gamma$) and (p,$\gamma$)
reactions from close to their respective driplines to stability, and for
other n-, p-, and $\alpha$-induced reactions with positive $Q$-values
near stability. A prescription on deriving rates for inverse reactions
with negative $Q$-values is given, as is a listing of the necessary
partition functions.

It should further be noted that only purely theoretical rates are given
here which do not use any direct experimental information (except for nuclear
masses and excited state information where available). The methods to predict
nuclear properties needed in the statistical model calculations are chosen
to be as reliable as possible in order to retain predictive power. This
is a compromise which may lead to locally enhanced inaccuracies but it
emphasizes the importance of reliable predictions of rates far off stability.

In real applications, these rates should be supplemented or replaced with
experimental rates as they become available. Such a combination of theoretical
and experimental rates is provided, e.g., in the REACLIB compilation. Latest
information on the current version of REACLIB can be found on the WWW at
\emph{\url|http://ie.lbl.gov/astro.html|}. Further details on the NON-SMOKER
code and the cross section and reaction rate calculations are presented
at \emph{\url|http://quasar.physik.unibas.ch/~tommy/reaclib.html|}. Rates
including further mass models can also be obtained from the authors on
request or directly at the latter URL.

\section*{Acknowledgements}

This work was supported in part by the Swiss National Science Foundation
(grant 2000-053798.98) and the Austrian Academy of Sciences (APART). 
T. R. is a PROFIL fellow of the Swiss National Science Foundation 
(grant 2124-055832.98). We thank S.E. Woosley and R.D. Hoffman for discussions.
We also want to thank the Consulting Editor Wilfried Scholz
for helpful comments and help in the
preparation of the final manuscript.

\newpage
\section{Explanation of Tables}

\subsection*{Table IA: Neutron-Induced Reaction Rates Available in Table II}

This is an overview of which neutron-induced reaction rates are available
in the printed and the online versions. The full lines delimit the range
of rates in the electronic version as given in Table A for the FRDM. 
The entries at
a single neutron and proton number specify the reactions on the given target
nucleus listed in the printed Table II. Only reactions with positive 
\( Q \)-value are shown. In addition to the marked rates,
their reverse rates (with
negative \( Q \)-value) can be inferred from the information in Table II
as explained in Sec.~\ref{sec:revpar}. The reactions are denoted as follows:
\begin{lyxlist}{00.00.0000}
\item [G](n,\( \gamma  \))
\item [P](n,p)
\item [A](n,\( \alpha  \))
\end{lyxlist}
The box at the lower left corner gives the location in the $Z$, $N$
plane of the final nucleus relative to the target nucleus for
(n,$\gamma$), (n,p), and (n,$\alpha$), thereby specifying also the
inverse reaction fits derivable from Table II.

\subsection*{Table IB: Proton-Induced Reaction Rates Available in Table III}

Same as Table IA but for proton-induced reactions. The marked reactions
correspond to the entries in Table III. The reactions are denoted as follows:
\begin{lyxlist}{00.00.0000}
\item [G](p,\( \gamma  \))
\item [N](p,n)
\item [A](p,\( \alpha  \))
\end{lyxlist}
The box at the lower left corner gives the location in the $Z$, $N$
plane of the final nucleus relative to the target nucleus for
(p,$\gamma$), (p,n), and (p,$\alpha$), thereby specifying also the
inverse reaction fits derivable from Table III.

\subsection*{Table IC: Alpha Particle-Induced Reaction Rates Available
in Table IV}

Same as Table IA but for \( \alpha  \)-particle induced reactions. The
marked reactions correspond to the entries in Table IV. The reactions are
denoted as follows:
\begin{lyxlist}{00.00.0000}
\item [G](\( \alpha  \),\( \gamma  \))
\item [N](\( \alpha  \),n)
\item [P](\( \alpha  \),p)
\end{lyxlist}
The box at the lower left corner gives the location in the $Z$, $N$
plane of the final nucleus relative to the target nucleus for
($\alpha$,$\gamma$), ($\alpha$,n), and ($\alpha$,n), thereby specifying also the
inverse reaction fits derivable from Table IV.

\subsection*{Table II: Neutron-Induced Reaction Rates}

Fits to stellar rates \( N_{A}\left\langle \sigma v\right\rangle ^{*} \)
for (n,\( \gamma  \)), (n,p), (n,\( \alpha  \)) reactions, calculated
including masses from the FRDM. The rates in cm\( ^{3} \) mole\( ^{-1} \)
s\( ^{-1} \) are computed by the use of Eq.~(\ref{fitpar}), with the temperature
given in units of 10\( ^{9} \) K. The fits are valid in the temperature
range \( T_{\mathrm{low}}^{\mathrm{fit}}<T_{9}\leq 10 \), 
with \( T_{\mathrm{low}}^{\mathrm{fit}} \)
given in the table. It should be noted that while the fit may still be
formally valid and accurate, the application of the statistical model may
not be justified at low temperatures. An estimate for the applicability
of the statistical model is given by \( T_{\mathrm{low}}^{\mathrm{HF}} \).
The following information is provided:

\begin{lyxlist}{00.00.0000}
\item [Target]Reaction target
\item [Reaction]Reaction type and final nucleus
\item [\( Q \)]Reaction \( Q \)-value
\item [\( J_{\mathrm{i}} \)]Target ground state spin (same as in Table V)
\item [\( J_{\mathrm{f}} \)]Final nucleus ground state spin (same as in Table V)
\item [\( T_{\mathrm{low}}^{\mathrm{HF}} \)]Estimate of the lower temperature
limit for the applicability of the (Hauser-Feshbach)
statistical model; ``n.c.'' indicates
that the limit was not calculated for the given reaction.
\item [\( T_{\mathrm{low}}^{\mathrm{fit}} \)]Lower temperature limit for the
fit; usually 0.01. Note that the fits give extrapolated rates below
$T_9=0.1$, which may be less accurate, especially if they are very
small.
\item [Dev]Fit accuracy \( \zeta  \) (Eq.~(\ref{eq:accur}))
\item [SEF]If the field is blank, the seven fit parameters below are followed
by the stellar enhancement factors (SEF) \( f^{*} \) (Eq.~(\ref{eq:sef}))
at the 16 temperatures given in the head of the table.\\
A value of 1 indicates that all SEF are unity; no SEF are printed.\\
A value of 0 indicates that no SEF were calculated.
\item [\( a_{0} \)\ldots{}\( a_{6} \)]Seven fit parameters for the forward
rate
\item [\( a_{0}^{\mathrm{rev}} \)]First parameter for the reverse rate fit (see
Sec.~\ref{sec:revpar})
\item [\( T_{9}^{\mathrm{SEF}} \)]Temperatures at which the SEF were calculated
\end{lyxlist}

\subsection*{Table III: Proton-Induced Reaction Rates}

Same as Table I for the reaction types (p,\( \gamma  \)), (p,n), 
(p,\( \alpha  \)).

\subsection*{Table IV: Alpha Particle-Induced Reaction Rates}

Same as Table I for the reaction types (\( \alpha  \),\( \gamma  \)),
(\( \alpha  \),n), (\( \alpha  \),p).

\subsection*{Table V: Partition functions}

Partition functions of isotopes for various temperatures calculated with
a level density making use of FRDM input. Included are only those partition
functions for nuclei involved in the reactions given in Tables II$-$IV.

\begin{lyxlist}{00.00.0000}
\item [Nuc]Isotope for which the partition functions are tabulated.
\item [$T_9$]Temperature (in 10$^9$ K) at which the partition functions
have been calculated.
\item [P]A value of 1 indicates that all partition function are unity; 
no partition functions are then printed explicitly.
\item [Spin] Ground state spin of nucleus, either from experiment \cite{audi95}
or from theory \cite{mnk97}.
\item [Partition~Functions:]
Partition functions normalized to the ground state (Eq.~(\ref{eq:partfunc}))
for the 20 temperatures specified in the table header.
\end{lyxlist}

\newpage

\begin{table}[ht]
\scriptsize
\caption{This table lists the isotope range of the full rate tables 
which are available electronically.
Given are the charge number \protect\( Z\protect \) of
a target and the lower and upper limits \protect\( N_{\mathrm{min}}\protect \)
and \protect\( N_{\mathrm{max}}\protect \) of the neutron number in the
isotopic chain.}

\begin{tabular}{rrrrrrrrrrr}
\hline\hline
 & \multicolumn{2}{c}{FRDM} & \multicolumn{2}{c}{ETFSI-Q}& & & 
\multicolumn{2}{c}{FRDM} & \multicolumn{2}{c}{ETFSI-Q} \\
$Z$ & $N_{\rm min}$ & $N_{\rm max}$ &$N_{\rm min}$ & $N_{\rm max}$ & 
\rule[-1.5pt]{16mm}{0mm} &
$Z$ & $N_{\rm min}$ & $N_{\rm max}$ &$N_{\rm min}$ & $N_{\rm max}$  \\
\hline\hline
  8 &   5 &  10 &     &     &&  47 &  41 & 113 &  41 & 111\\
  9 &   5 &  28 &     &     &&  48 &  42 & 115 &  42 & 112\\
 10 &   5 &  31 &     &     &&  49 &  43 & 117 &  43 & 113\\
 11 &   6 &  33 &     &     &&  50 &  44 & 119 &  44 & 114\\
 12 &   7 &  35 &     &     &&  51 &  46 & 121 &  46 & 115\\
 13 &   8 &  38 &     &     &&  52 &  47 & 124 &  47 & 124\\
 14 &   8 &  40 &     &     &&  53 &  48 & 126 &  48 & 126\\
 15 &   8 &  42 &     &     &&  54 &  49 & 128 &  49 & 128\\
 16 &   8 &  44 &     &     &&  55 &  51 & 130 &  51 & 130\\
 17 &   9 &  46 &     &     &&  56 &  52 & 133 &  52 & 132\\
 18 &   9 &  49 &     &     &&  57 &  53 & 135 &  53 & 133\\
 19 &  10 &  51 &     &     &&  58 &  55 & 137 &  55 & 134\\
 20 &  10 &  53 &     &     &&  59 &  56 & 139 &  56 & 135\\
 21 &  11 &  55 &     &     &&  60 &  58 & 141 &  58 & 136\\
 22 &  12 &  58 &     &     &&  61 &  59 & 144 &  59 & 137\\
 23 &  13 &  60 &     &     &&  62 &  61 & 146 &  61 & 138\\
 24 &  14 &  62 &  18 &  62 &&  63 &  62 & 148 &  62 & 139\\
 25 &  15 &  64 &  18 &  64 &&  64 &  64 & 150 &  64 & 150\\
 26 &  16 &  66 &  19 &  66 &&  65 &  65 & 153 &  65 & 152\\
 27 &  17 &  69 &  19 &  67 &&  66 &  67 & 155 &  67 & 154\\
 28 &  18 &  71 &  20 &  68 &&  67 &  69 & 157 &  69 & 155\\
 29 &  19 &  73 &  21 &  69 &&  68 &  70 & 159 &  70 & 156\\
 30 &  21 &  75 &  22 &  70 &&  69 &  72 & 161 &  72 & 157\\
 31 &  22 &  77 &  23 &  71 &&  70 &  73 & 164 &  73 & 158\\
 32 &  23 &  80 &  24 &  72 &&  71 &  75 & 166 &  75 & 159\\
 33 &  24 &  82 &  25 &  73 &&  72 &  77 & 168 &  77 & 160\\
 34 &  25 &  84 &  26 &  84 &&  73 &  78 & 170 &  78 & 161\\
 35 &  26 &  86 &  27 &  86 &&  74 &  80 & 173 &  80 & 162\\
 36 &  27 &  88 &  28 &  88 &&  75 &  81 & 175 &  81 & 163\\
 37 &  29 &  91 &  29 &  89 &&  76 &  83 & 177 &  83 & 177\\
 38 &  30 &  93 &  30 &  90 &&  77 &  85 & 179 &  85 & 179\\
 39 &  31 &  95 &  31 &  91 &&  78 &  87 & 182 &  87 & 182\\
 40 &  32 &  97 &  32 &  97 &&  79 &  88 & 184 &  88 & 184\\
 41 &  33 &  99 &  33 &  99 &&  80 &  90 & 186 &  90 & 186\\
 42 &  35 & 102 &  35 & 102 &&  81 &  92 & 188 &  92 & 188\\
 43 &  36 & 104 &  36 & 104 &&  82 &  93 & 191 &  93 & 191\\
 44 &  37 & 106 &  37 & 106 &&  83 &  95 & 193 &  95 & 193\\
 45 &  38 & 108 &  38 & 108 &&  84 &  98 & 193 &  98 & 193\\
 46 &  40 & 110 &  40 & 110 &&  85 & 101 & 195 & 101 & 195\\
\hline\hline
\end{tabular}
\end{table}

\newpage
\section*{The remaining tables can be found on the ADNDT server}
\noindent
or at
\url{http://quasar.physik.unibas.ch/~tommy/adndt.html}.

\end{document}